\documentclass[useAMS,usenatbib,usegraphicx]{mn2e}
   \title[Substructure in a wind]{Hydrodynamical adaptive mesh refinement simulations of turbulent flows - I. Substructure in a wind}

   \author[L.~Iapichino et al.]{L.~Iapichino,\thanks{E-mail:luigi@astro.uni-wuerzburg.de} J.~Adamek, W.~Schmidt and J.C.~Niemeyer \\
Institut f\"ur Theoretische Physik und Astrophysik, Universit\"at W\"urzburg, Am Hubland, D-97074 W\"urzburg, Germany}

\begin{document}

\date{Accepted 2008 February 5. Received 2008 January 29; in original form 2007 September 18}

\pagerange{\pageref{firstpage}--\pageref{lastpage}} \pubyear{2008}

\maketitle

\label{firstpage}

\begin{abstract}
The problem of the resolution of turbulent flows in adaptive mesh
refinement (AMR) simulations is investigated by means of 3D
hydrodynamical simulations in an idealised setup, representing a
moving subcluster during a merger event. 
AMR simulations performed with the usual refinement criteria based on
local gradients of selected variables do not properly resolve the
production of turbulence downstream of the cluster. Therefore we
apply novel AMR criteria which are optimised to follow the evolution
of a turbulent flow. 
We demonstrate that these criteria provide a better resolution of the
flow past the subcluster, allowing us to follow the onset of the shear
instability, the evolution of the 
turbulent wake and the subsequent back-reaction on the subcluster core
morphology.  
We discuss some implications for the modelling of cluster cold fronts.
\end{abstract}

\begin{keywords}
Hydrodynamics -- Instabilities -- Methods: numerical -- Galaxies: clusters: general -- Turbulence
\end{keywords}

\section{Introduction}
\label{intro}

The role of turbulence in astrophysics and, in particular, in
cosmic structure formation has been attracting increasing
attention, along with observational and theoretical indications of its
relevance. There are different observational hints converging towards
the turbulent character of the intra-cluster medium (ICM)
(\citealt{sfm04,cfj04,ev06,rcb05,rcb06}, but see also
\citealt{fsc03}). From an observational point of view, the next
generation of X-ray observatories will be able to observe the Doppler
shifts of emission lines of heavy ions due to bulk motions in the ICM,
thus providing information about the turbulent state of the
flow. Numerical studies predict that gas bulk motions and turbulence
in the ICM can be a consequence of cluster merging
\citep{r98,nb99,t00,rs01,dvb05}. In their model, \citet{ssh06}
identify three physical regimes for turbulence production and decay in
clusters, the latest one being dominated by turbulent production in
the wakes of minor mergers. This phase would play a key role also for
the magnetic field amplification in the ICM. 

Merging is a crucial process in cosmic structure formation. It is
predicted in the framework of hierarchical clustering and can be studied
theoretically by means of numerical simulations
(e.g.~\citealt{rbl96,rsm98,mbb07}). Observationally, modern X-ray
observatories are able to detect substructures in the ICM that are
indicative of merging \citep{snb03}. Among the 
first results of galaxy cluster observations with {\it Chandra} was the
discovery of ``cold fronts'' in the merging clusters A2142 and A3667,
and later in many other clusters (see \citealt{mv07} for a
review). The origin of cold fronts in merger clusters has been
ascribed to infalling substructures, although the cold fronts in the
cores of cooling core clusters have a different interpretation
\citep{th05,am06}. 

Cosmological simulations of galaxy cluster formation have confirmed
the merging scenario for the transient formation of cold fronts
\citep{bem02,nk03}. The problem has also been addressed with
simplified setups, which allow a better control over the involved
physical parameters, in 2D \citep{asp03,hcf03,afm04,xcd07} and 3D
simulations \citep{t05a,t05b,afm05,afm07,dp07}. 

The theoretical study of turbulence in the framework of full
cosmological simulations is a challenging task and presents the
typical problems of numerical simulations of strongly clumped
media. Adaptive mesh refinement (AMR) is a viable tool for saving
computational resources and modelling the large dynamic range in a
proper way \citep{n04}. The choice of the most suitable mesh
refinement criterion is thus a delicate compromise between following
the flow structure in the most accurate way and exploiting the
advantage of saving computational time and memory. 

This study (Paper I) is the first part of a broader project on the
physics of galaxy clusters, focused on investigating the
generation of turbulence in the ICM by means of AMR simulations. In
Paper I, we present new AMR criteria that are
more suitable for resolving the production of turbulence than those
commonly used in cosmological simulations. They were tested in a
simple setup, reminiscent of a wind tube experiment, but designed in
order to be representative of an idealised subcluster merger. This
test problem is a useful and controlled testbed for the new tools, far from
the complexity of a more realistic cosmological
simulation. Nevertheless, it is similar to previous studies
of subcluster mergers and can provide useful insights about
the physics of the minor mergers, relevant for the generation of
turbulence in the ICM. In contrast with earlier work, we focus
our analysis on the turbulent wake of the subcluster rather than on
the cold front, showing the difference of the evolution of the
Kelvin-Helmholtz instability (KHI) resulting from the use of different
AMR criteria. 

The second step, described in \citet{in08} (hereafter Paper II), is to apply the same tools to simulations of structure 
formation and to study the turbulent flow in the ICM. The
results of the present work will thus be compared with the outcomes of
more realistic simulations of minor mergers.  

The paper is structured as follows: the new refinement criteria are
defined in in Sec.~\ref{simulations}. The numerical setup of the
performed simulations and the features of the KHI are introduced in
Sects.~\ref{setup} and \ref{theory}, respectively. The results are
presented in Sec.~\ref{results} and discussed in
Sec.~\ref{discussion}.

\section{Refinement criteria and resolution of turbulent flows}
\label{simulations}

Three different AMR criteria are used in the simulations discussed in
this work. 
A rather general and widely used\footnote{This is the criterion used in
  public code {\sc enzo}, employed for this work (Sec.~\ref{setup}).} criterion
is based on the local gradients of all thermodynamical variables.
If the relative slope $|(q(i+1)-q(i-1))/q(i)|$ of a variable $q$ in a
cell $i$ is larger than a given threshold, that cell is marked for
refinement. Numerical tests showed that, for our subcluster problem,
this criterion quickly leads to a high level of refinement in the entire
computational domain. This shortcoming was fixed by allowing the
refinement on the local gradients of only selected variables. The AMR
criterion named ``1'' in our work is based on the local gradients of
density and internal energy, with thresholds $0.24$ and $0.25$
respectively, tuned for optimal performance. 

Two novel AMR criteria, developed for tracking the evolution of a turbulent
flow \citep{sfh08}, were tested
in this study. In both cases, the control variables for refinement are
given by scalars probing small-scale features of the flow. One example is the 
modulus of the vorticity vector $\bmath{\omega} = \nabla \times \bmath{v}$ (the
curl of the velocity field) that is expected to become high in
regions filled by turbulent eddies. In addition, the mechanism for
triggering refinement has been modified. Rather than normalising the
control variables 
in terms of characteristic quantities and comparing to prescribed
threshold values, the new criteria use regional thresholds for triggering
refinement which are based on a comparison of the cell value of the
variable $q(\bmath{x},t)$ with the average and the standard deviation of $q$,
calculated on a local grid patch: 
 \begin{equation}
\label{local}
q(\bmath{x},t) \ge \langle q \rangle_i(t) + \alpha \lambda_i(t)
\end{equation}
where $\lambda_i$ is the maximum between the average $\langle q
\rangle $ and the standard deviation of $q$ in the grid patch $i$, and $\alpha$
is a tunable parameter. 
This technique can easily handle highly inhomogeneous
problems such as subcluster mergers without a priori knowledge of
the flow properties.

We define criterion ``2'' using the square of the vorticity,
$\omega^2$, as the control variable. Refining upon criterion ``2'' only,
one can expect to resolve turbulent eddies which are associated with
high vorticity but not flow features arising from pure compression
such as shock fronts. For this reason, in criterion 
``3'', the control variable is given by the rate of compression, i.e., the
negative time derivative of the divergence $d=\nabla\cdot\bmath{v}$.
In combination with refinement by vorticity, this criterion is
intended to refine flow regions in which steep density gradients are
developing. One could also refine by $-d$ or $d^{2}$ in order to
capture compression effects. However, this would give rise to full
refinement of highly compressed gas, despite the fact that
the interiors of compressed regions do not necessarily exhibit rich
small-scale structure. Maxima of the rate of compression, on the
other hand, are typically found at the interfaces between dense and
rarefied gas. As we have seen in applications, a drawback of
criterion ``3'' is that refinement can be triggered prematurely in
some instances. This undesired side-effect can be suppressed by
introducing a lower cutoff for the refinement thresholds.

\section{Numerical simulations}
\label{setup}

\begin{table}
\caption{Scheme of the performed simulations.}
\label{list}
\centering
\begin{tabular}{c| c c}
\hline
Simulation & number of AMR levels & AMR criterion \\
\hline
$A$ & 5 & 1 \\
$B$ & 4 & 1 \\
$C$ & 6 & 1 \\
$D$ & 5 & 2 \\
$E$ & 5 & 2+3 \\
\hline
\end{tabular}
\end{table}

The list of the performed calculations is reported in Table
\ref{list}. In the first column, the simulations are identified by a
letter. The second column reports the number $n$ of AMR levels which
are allowed in the simulation.  The root grid resolution of the runs
is only $16^3$ but the effective resolution is finer because AMR is
used. The refinement factor is 2, thus the effective resolution is
$(16 \times 2^n)^3$\ grid cells. In the standard case of five
additional AMR levels, the effective resolution is therefore $(16
\times 2^5)^3 = 512^3$ grid elements. 

The third column of Table \ref{list} indicates which criterion was
used for performing the grid refinement, according to the 
definitions of Sec.~\ref{simulations}. The simulations $A$, $B$ and
$C$ make use of criterion ``1'' in a resolution study, whereas the
simulations $D$ and $E$ were performed with the novel AMR criteria
``2'' with threshold $\alpha_2 = 0.2$, and ``$2 + 3$'' with $\alpha_2
= 1.0$, $\alpha_3 = 2.0$, respectively. 

The 3D simulations were performed using the AMR, grid-based hybrid
(hydrodynamical plus N-Body) code {\sc enzo} \citep{obb05}\footnote{{\sc enzo}
  homepage: http://lca.ucsd.edu/portal/software/enzo}, with the N-Body
section of the code switched off. The
primary hydrodynamic method used in {\sc enzo} is based on the piecewise
parabolic method (PPM) \citep{wc84}, modified for the study of
cosmology \citep{bns95}. Also available is the hydrodynamic
finite-difference algorithm of the {\sc zeus} code \citep{sn92a,sn92b}, used
for tests in Sec.~\ref{zeus}.  

Since we assumed ideal adiabatic physics for our calculations, the
simulations are scale-free and all quantities were rescaled to get
numbers of the order unity, as shown in \citet{rdr04}. However, for
the sake of clarity, all quantities are reported in CGS units. An
adiabatic equation of state was used, with $\gamma = 5/3$.  

In the initial conditions, an isothermal and spherical symmetric
object is set in the computational domain according to a beta density
profile \citep{cff78}: 
\begin{equation}
\label{beta-profile}
\rho(r) = \rho_\rmn{c} \left[1 + \left( \frac{r}{r_\rmn{c}} \right)^2
\right]^{-3 \beta / 2} \,\,,
\end{equation}
where $\rho$ is the gas density, $r$ is the distance from the centre,
$\rho_\rmn{c}$ is the central density, and $r_\rmn{c}$ is the core
radius. The most important difference between this setup and a typical
wind tube experiment (cf.~\citealt{mwb93,vam97,ams06}) is the presence
of gravity. The gravitational acceleration is provided by a
fixed spherical dark matter distribution following a King
profile,
\begin{equation}
\label{dm-profile}
\rho_\rmn{DM}(r) = \rho_\rmn{DM,c} \left[1 + \left(
    \frac{r}{r_\rmn{c}} \right)^2 \right]^{-3 / 2}\,\,, 
\end{equation}
with $\rho_\rmn{DM,c} = 10\ \rho_\rmn{c}$.  A cutoff was set at $r =
5\ r_\rmn{c}$, both for the dark matter density profile and the
gravitational acceleration \citep[cf.~][]{t05a}. Given the King
profile for the dark matter component of an isothermal cluster, the
beta profile is the hydrostatic equilibrium solution for the gas
density. The hydrostatic equilibrium is enforced on the discrete
computational grid by the relaxation procedure of the initial state
described by \citet{zdz02}. 

The interaction of the subcluster with a background wind mimics the
merger process in a simplified manner. A uniform velocity field along
the $x$-axis is imposed in the background medium. 
In this simplified setup, we implicitly assume that the merging process has a
negligible impact on the dark matter density profile of the
subcluster. 

The numerical values of the parameters defined above were chosen as
typical of a subcluster merger. 
 We set $\rho_\rmn{c} = 6.3\times 10^{-27}\ \rmn{g\ cm^{-3}}$,
 $r_\rmn{c} = 7.7\times 10^{23}\ \rmn{cm} = 250\ \rmn{kpc}$ and $\beta
 = 0.6$, which result in a subcluster temperature $kT = 3.65\
 \rmn{keV}$. The background medium has constant values of
 density $\rho_\rmn{b} = 7.9\times 10^{-28}\ \rmn{g\ cm^{-3}}$ and
 temperature $kT_\rmn{b} = 8.0\ \rmn{keV}$. The background velocity is
 chosen as $v_\rmn{b} = 1.6\times 10^3\ \rmn{km\ s^{-1}}$,
 $v_\rmn{b}/c_\rmn{b} = 1.1$, where $c_\rmn{b}$ is the sound speed of
 the background. The interface between the subcluster and the uniform
 medium is defined at the locations where the subcluster and background
 gas pressures are equal. 

The computational domain has a size of $[0, 4\ \rmn{Mpc}]^3$, meaning
that the effective resolution of $512^3$ corresponds to an effective
spatial resolution of $7.8\ \rmn{kpc}$. The boundary conditions are
inflowing at the left $x$-boundary and outflowing elsewhere. The
subcluster is initially resolved with the finest AMR level available,
in order to minimise any mapping errors related to the hydrostatic
equilibrium of the initial state.

\section{The Kelvin-Helmholtz instability}
\label{theory}

The simulations are carried out in the rest frame of the subcluster,
set at rest in a moving background medium. The outer subcluster
interface is thus subject to the Kelvin-Helmholtz instability. If
gravity is neglected, the KHI may grow for all values of wavenumber
$k$. The linear growth timescale of the KHI, in the case of a flat
interface between two incompressible fluids of different densities, is
(\citealt{dr04}; cf.~\citealt{ams06}): 
\begin{equation}
\label{khigrowth}
\tau_\rmn{KH} = \frac{2 \pi (\rho_\rmn{1} + \rho_\rmn{2})}{k (\rho_\rmn{1} \rho_\rmn{2})^{1/2} V}
\end{equation}

The geometry of the subcluster is different from this
idealised case, but an order-of-magnitude estimate of $\tau_\rmn{KH}$
can be obtained by setting $\rho_\rmn{1} = 0.3\, \rho_\rmn{c}$
(subcluster density at the interface with the background medium),
$\rho_\rmn{2} = \rho_\rmn{b}$, $k = 2 \pi / R$, where $R \sim 1.7\,
r_\rmn{c}$ is the distance from the subcluster centre to the interface
with the background medium. The velocity $V$ is smaller than
$v_\rmn{b}$ because a bow shock forms in front of the subcluster and
$V$ has to be determined from the post-shock conditions
(cf.~\citealt{ams06}). For this rough estimate it is adequate to
assume $V \approx 0.7\, v_\rmn{b}$ (also in agreement with the
simulations). With these choices, $\tau_\rmn{KH} \approx 0.8\
\rmn{Gyr}$. This is an estimate of the time required to the KHI to
perturb the spherical shape of the subcluster substantially. We set
the total simulation time to $4\ \rmn{Gyr}$. 

The previous timescale analysis motivates the choice of neglecting
radiative cooling in the performed simulations. The effect of cooling
is not expected to be prominent because both for the subcluster and
the background medium the cooling timescale is about two times longer
than the total simulation time, and much longer than $\tau_\rmn{KH}$. 

The presence of the gravitational acceleration stabilises a
perturbation of wavenumber $k$ if (\citealt{p84}; cf.~\citealt{mwb93}) 
\begin{equation}
\label{khigravity}
g \ga \frac{\rho_\rmn{1} \rho_\rmn{2} V^2 k}{\rho_\rmn{1}^2 - \rho_\rmn{2}^2}
\end{equation}
where $g$ is the modulus of the gravitational acceleration at the
interface. In the subcluster case, this stabilisation only affects
wave numbers that are much smaller than $2 \pi / R$. Gravity thus
fails to stabilise the subcluster against the KHI. 

The loss of gas from the subcluster atmosphere due to the ram pressure
of the background flow, discussed by \citet{hcf03}, is not effective
for our choice of parameters. The gas stripping and its subsequent
mixing into the ambient medium is instead caused by the KHI, and will
be described in Sect.~\ref{amr}.

\section{Results}
\label{results}

\subsection{Evolution of the subcluster}
\label{evolution}

\begin{figure*}
\centering
  \includegraphics[width=17cm]{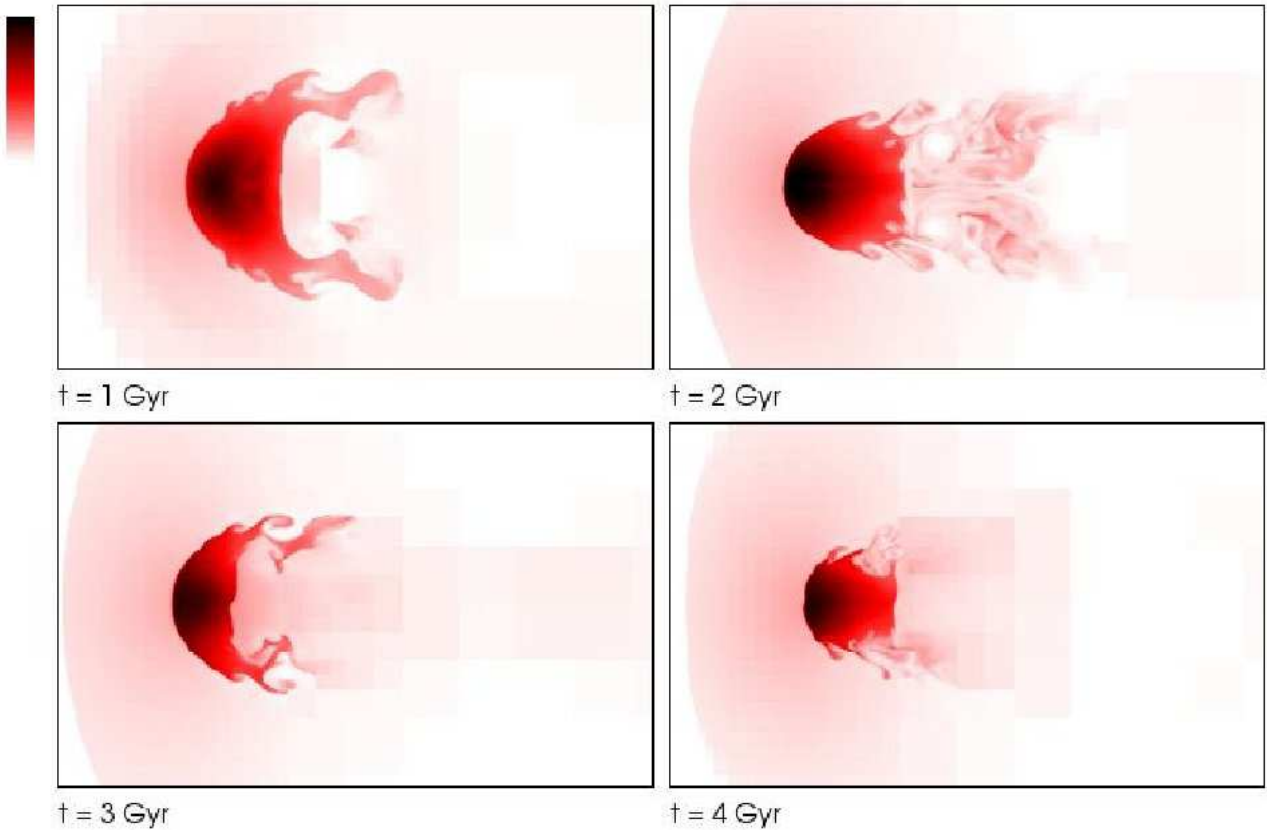}
  \caption{Density slices ($xz$ plane) of a part of the computational
    domain ($2.6 \times 1.6\ \rmn{Mpc}$), showing the subcluster
    evolution at different times, for the run $A$. The density is
    linearly colour coded, following the colour bar on the upper
    left. Time is indicated at the lower left of each panel.} 
  \label{figure-evolution}
\end{figure*}

The morphological evolution of the idealised subcluster is shown in
the four density slices of Fig.~\ref{figure-evolution}. The figure
refers to run $A$, which (as discussed in Sec.~\ref{resolution}) has
a suitable spatial resolution for our problem. 

At $t = 1\ \rmn{Gyr}$, a bow shock has formed in front of the
subcluster and the KHI is growing at the sides. At $t = 2\ \rmn{Gyr}$
a turbulent, eddy-like flow in the subcluster wake is clearly
visible. Yet it is not resolved at later times, as will be discussed
in Sec.~\ref{amr}. The KHI leads to mass stripping in the
subcluster, which will be quantified in Sect.~\ref{resolution}. 

\begin{figure}
  \resizebox{\hsize}{!}{\includegraphics{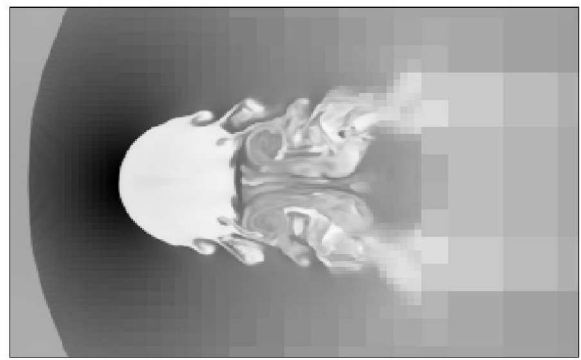}}
  \caption{Temperature slice ($xz$ plane) of the subcluster at $t = 2\
    \rmn{Gyr}$ for the run $A$ (AMR on gradients of density and
    internal energy). The temperature is linearly colour coded in
    gray-scale.} 
  \label{temp-slice}
\end{figure}

From $t = 2\ \rmn{Gyr}$, the morphology which is visible at the left
interface of the subcluster (Fig.~\ref{temp-slice}) closely resembles
a typical cold front structure. Based on {\it Chandra} observations of
A3667, \citet{vmm01a,vmm01b} and \citet{vm02} state that the very
existence of the cold fronts requires the suppression of the KHI and
of the thermal conduction at the interface between the subcluster gas
and the ICM, likely caused by the presence magnetic fields. The MHD
simulations of \citet{afm04}, \citet{afm05} and \citet{afm07} include
anisotropic thermal conduction and show an effective stabilisation of
the cold front. The suppression of the thermal conduction is needed to
reproduce the observed temperature profiles of the cold front in A3667
in the SPH simulations of \citet{xcd07}. 

Our simulations include neither heat conduction nor magnetic
fields. Similar to \citet{hcf03}, a ``cold front-like''  structure is
thus visible but, according to the above cited simulations, it does
not catch the physics of the astrophysical problem. However, its
analysis is not the primary goal of this work. In
Sec.~\ref{discussion}, we will extend the discussion of the subcluster
evolution to examples which take into account such additional
physics. 

\subsection{Resolution study}
\label{resolution}

\begin{figure*}
\centering
  \includegraphics[width=17cm]{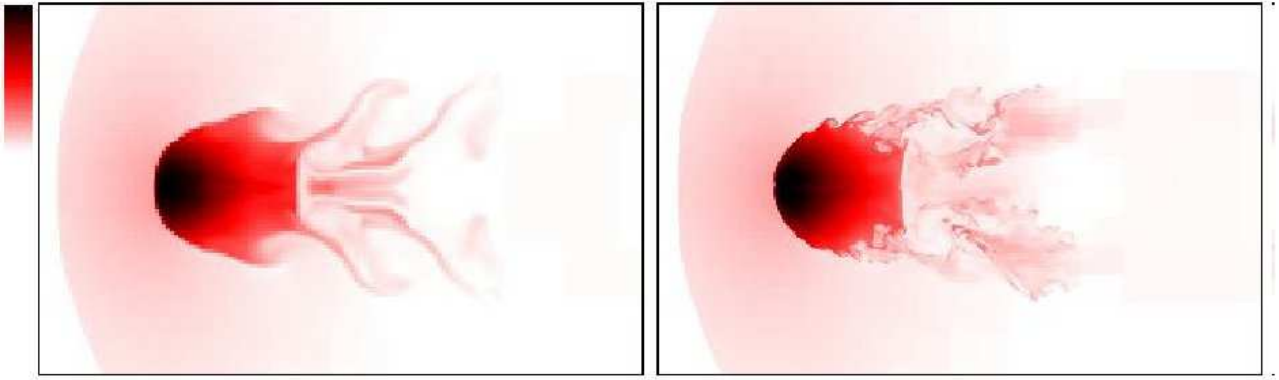}
  \caption{Density slices of the subcluster at $t = 2\ \rmn{Gyr}$
    (cf.~Fig.~\ref{figure-evolution}, upper right panel), in
    simulations with different effective resolution. Left: run
    $B$. Right: run $C$. The density is linearly colour coded,
    according to the colour bar on the left.} 
  \label{resolution-density}
\end{figure*}

A comparison of AMR runs with different resolution is useful for
understanding how effectively (or not) the idealised subcluster is
modelled in the simulations. 
With this intent, besides the presented run $A$ we performed the
additional runs $B$ and $C$ (cf.~Table \ref{list}), which implement
the same AMR criterion as $A$ and have the same root grid resolution,
but allow different numbers of additional AMR levels. Figure
\ref{resolution-density} permits a morphological comparison at $t = 2\
\rmn{Gyr}$. The run $B$ allows four additional levels of refinement
and has a lower effective resolution than run $A$. Comparing
Fig.~\ref{resolution-density}, left, and the upper right panel of
Fig.~\ref{figure-evolution}, a less developed shear instability and an
almost regular pattern in the flow at the subcluster tail  is
noticeable in the former. The crucial features of the subcluster
evolution are therefore not properly followed at this level of
resolution. On the other hand, the analysis of more resolved run $C$
(six additional AMR levels) shows obviously a better resolution than
run $A$ and an enhanced development of the KHI at the sides of the
subcluster. Nevertheless, it does not provide any new element for the
study of the problem, and therefore the effective resolution of the
run $A$ is considered to be adequate for our investigation.  

\begin{figure}
  \resizebox{\hsize}{!}{\includegraphics{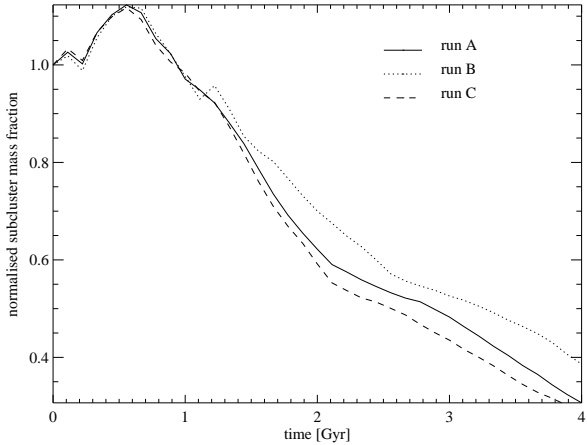}}
  \caption{Temporal evolution of subcluster baryonic mass fraction for
    the runs $A$ (solid line), $B$ (dotted line) and $C$ (dashed
    line). The mass fraction is normalised to its value at $t = 0\
    \rmn{Gyr}$.} 
  \label{cloud}
\end{figure}

A more quantitative comparison is obtained by defining a diagnostic
of gas stripping resulting from the development of the
KHI. A suitable quantity is adapted from the ``cloud mass fraction''
defined by \citet{ams06}, hence we consider the gas with $T < 0.9\
T_{\rmn{b}}$ and $\rho > 0.32\ \rho_{\rmn{c}}$ as belonging to the
subcluster. After a transient phase related to the development of the
bow shock, where the ``subcluster mass'' seems to increase, this
quantity (Fig.~\ref{cloud}) decreases with time. Similar to
\citet{ams06}, the subcluster stripping is more pronounced in the more
resolved runs, due to the better resolved small scale instability and mixing.

\subsection{AMR study}
\label{amr}

One of the most interesting features of the subcluster evolution in
run $A$ is the loss of resolution in the wake at late times. This is
an effect of the AMR criterion chosen in that simulation, which cannot
effectively refine the turbulent tail (or, at least, without wasting
the advantage of AMR, with unnecessary refinement in almost the entire
computational domain). 

\begin{figure}
  \resizebox{\hsize}{!}{\includegraphics{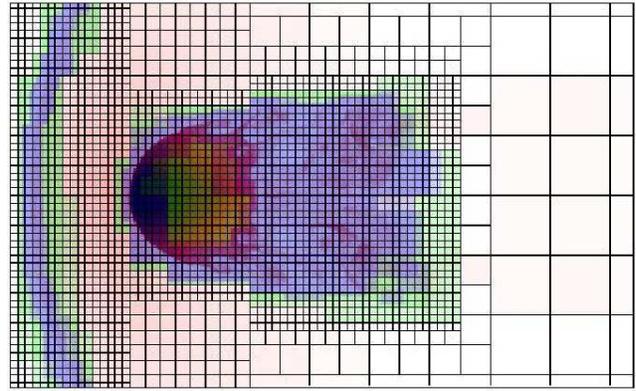}}
  \caption{Density slice as in Fig.~\ref{figure-evolution}, upper
    right panel, with the mesh structure superimposed. Grids of AMR
    levels from 0 to 3 are rendered as mesh structures, whereas for
    ease of visualisation grids of level 4 and 5 are only rendered
    with colours green and blue, respectively.} 
  \label{density-mesh-a}
\end{figure}

In a study of cold front physics, one could think that the
refinement of run $A$ (Fig.~\ref{density-mesh-a}) is acceptable,
because the attention is focused on the better resolved part of the
computational domain. We dispute this statement in this
section. To this aim, we checked whether alternative AMR
criteria can better track the turbulent wake, provided the rest of the
cluster morphology is also appropriately refined. 

\begin{figure}
  \resizebox{\hsize}{!}{\includegraphics{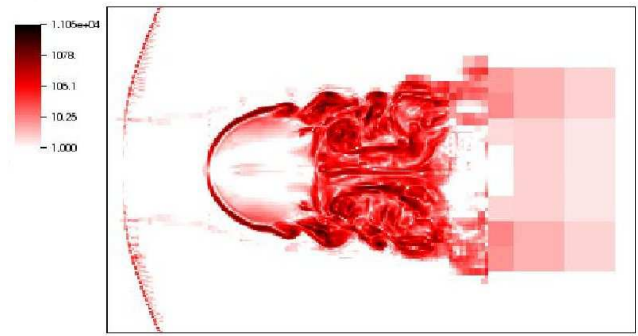}}
  \caption{Slice of the subcluster at $t = 2\ \rmn{Gyr}$ for the run
    $A$ (cf.~Fig.~\ref{figure-evolution}, upper right panel), showing
    the square of the vorticity modulus $\omega^2$. The quantity is
    colour coded on a logarithmic scale in code units.} 
  \label{vorticity-a}
\end{figure}

Velocity fluctuations at all scales are the prominent feature of a
turbulent flow. It suggests that quantities related to the spatial
derivatives of velocity can be particularly suitable for the
characterisation of the flow. Figure \ref{vorticity-a}, which shows a
slice of the square of the vorticity modulus in run $A$, clearly
confirms this idea: the morphology of the turbulent flow is well
tracked in comparison with the less contrasted density and
temperature eddies, which are not very suitable for triggering the
mesh refinement. 

\begin{figure*}
\centering
  \includegraphics[width=17cm]{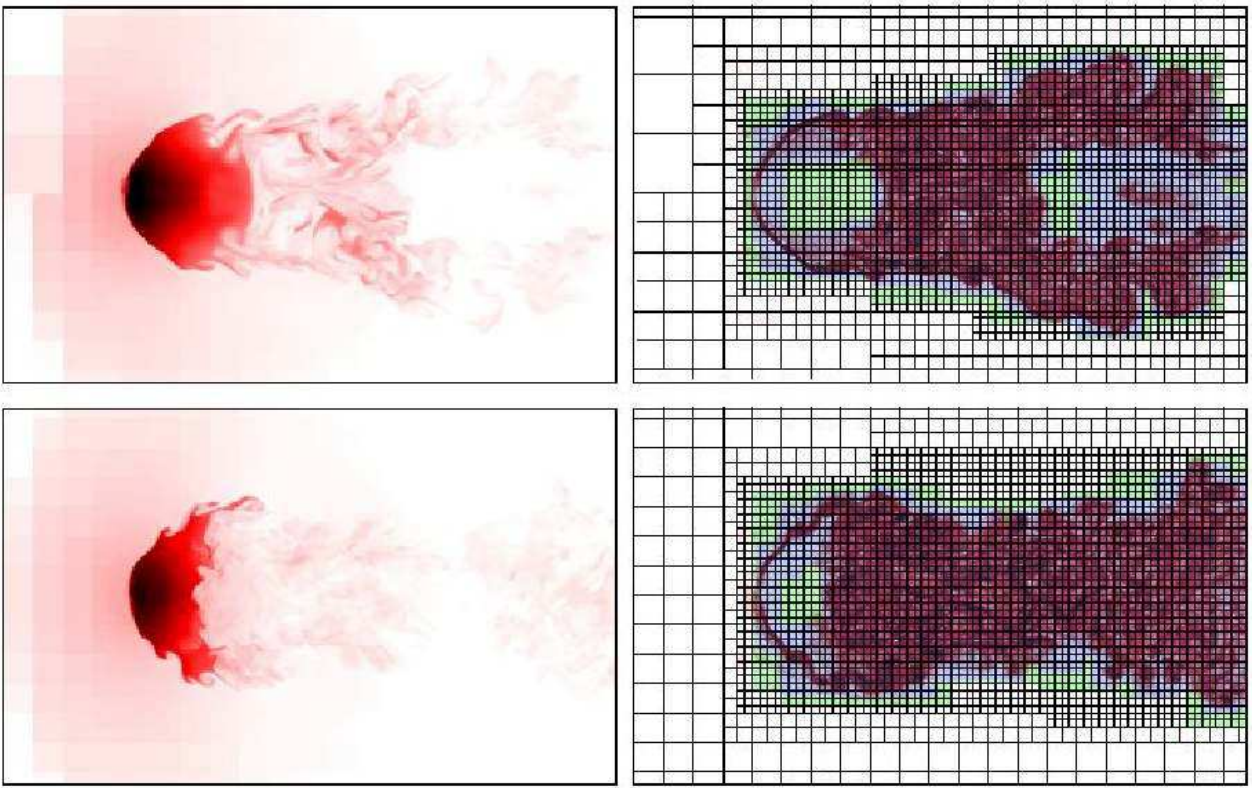}
  \caption{Slices ($xz$ plane) for the run $D$, showing density (left)
    and square of the vorticity modulus, with the mesh superimposed
    and visualised according to the rendering used in
    Fig.~\ref{density-mesh-a} (right). Upper panels: slices $t = 2\
    \rmn{Gyr}$. Lower panels: slices $t = 3\ \rmn{Gyr}$.} 
  \label{run-d}
\end{figure*}

\begin{figure*}
\centering
  \includegraphics[width=17cm]{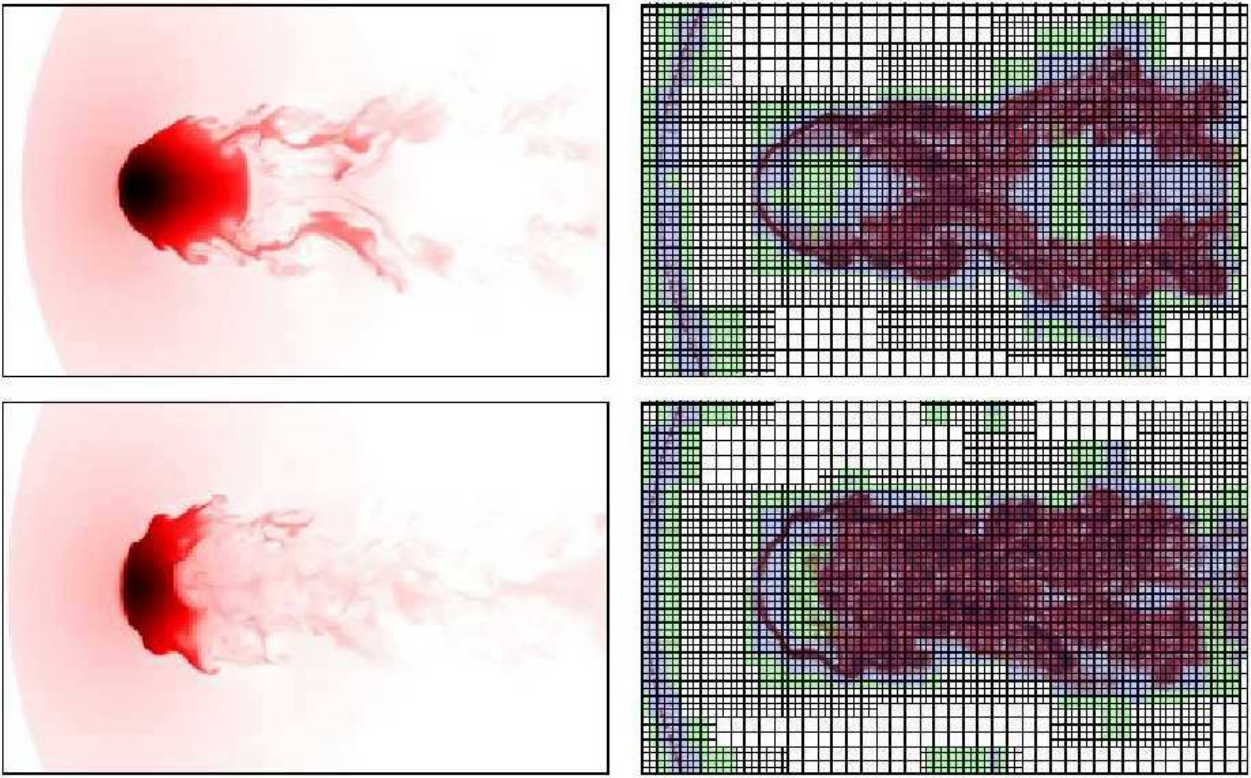}
  \caption{Same as Fig.~\ref{run-d}, for run $E$.}
  \label{run-e}
\end{figure*}

The refinement criteria ``2'' and ``3'', presented in
Sect.~\ref{simulations}, are based on this concept. 
Their effect on the resolution of the turbulent subcluster wake can be
evaluated by the analysis of the runs $D$ and $E$ (Figs.~\ref{run-d}
and \ref{run-e}, respectively, at  $t = 2$ and $3\ \rmn{Gyr}$). From
the density and vorticity slices one can immediately recognise that,
in both simulations, the subcluster wake is effectively resolved down
to the finest available AMR levels. Moving out of the subcluster, the
refinement level decreases gradually. 

A closer look reveals that there are some differences between the two
simulations. As anticipated in Sec.~\ref{simulations}, the refinement
criterion ``2'' alone is unable to resolve  the bow shock in front of the
subcluster at the finest levels of 
refinement (run $D$). It is difficult to state whether this has any effect on
the development of the KHI, but it is certainly not desirable to
underresolve such an important feature of the problem, well resolved
in run $E$.  

The subcluster tail in run $E$ at $t = 2\ \rmn{Gyr}$
(Fig.~\ref{run-e}, upper left) has a more filamentary pattern,
intriguingly similar to the more resolved run $C$
(cf.~Fig.~\ref{resolution-density}, right). 

At $t = 3\ \rmn{Gyr}$, in both runs $D$ and $E$ the subcluster appears
more perturbed and prone to the KHI than run $A$
(Fig.~\ref{figure-evolution}, lower left panel). Since the subcluster
front is well resolved with the refinement criteria used in all
simulations, this difference is (at least partly) to be ascribed to
the back-reaction of the tail. The turbulent eddies in the subcluster
wake are better resolved in runs $D$ and $E$, and partly disturb the
morphology of the subcluster core even though it is well
resolved. This effect of back-flow is rather similar to that described
by \citet{hcf03}, which identify the displacement of the subcluster
core with respect to the potential well as an additional source. 

\begin{figure}
  \resizebox{\hsize}{!}{\includegraphics{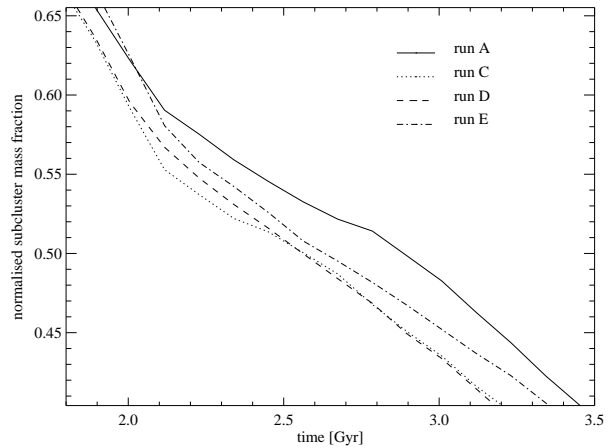}}
  \caption{Detail of the temporal evolution of normalised subcluster
    mass fraction for the runs $A$ (solid line), $C$ (dotted line),
    $D$ (dashed line) and $E$ (dot-dashed line).} 
  \label{cmf-compare}
\end{figure}

As a measure of the action of the KHI, the subcluster mass has been
introduced in Sec.~\ref{resolution}. It is useful to investigate
quantitatively the effect of the back-flow, comparing the time
evolution of this quantity in the runs $A$, $C$, $D$, and $E$
(Fig.\ref{cmf-compare}). Interestingly, after $t = 2\ \rmn{Gyr}$ the
stripping for the runs $D$ and $E$ is more effective than for the run
$A$, and follows more closely the curve of the more resolved run
$C$. The seeming analogy comes from the convergence of two different
features of the small-scale mixing, which for run $C$ is more active
at the sides of the subcluster, whereas for the runs $D$ and $E$ is
triggered mainly by the tail back-flow. 

\begin{table}
\caption{Mass-weighted rms velocity in the test volume of $(400\
  \rmn{kpc})^3$, at  $t = 3\ \rmn{Gyr}$. The run labelled as
  ``Static'' will be introduced in Sect.~\ref{static}.} 
\label{rms}
\centering
\begin{tabular}{c|c}

\hline
run & $v_{\rmn{rms}}\ (\rmn{km\ s^{-1}})$ \\
\hline
$A$    &  260 \\
$D$    &  530 \\
$E$    &  530 \\
Static &  620 \\
\hline
\end{tabular}
\end{table}

In order to further quantify the effectiveness of the tested AMR
criteria, the mass-weighted root mean squared (rms) velocity
$v_{\rmn{rms}}$ is calculated in a test volume of $(400\
\rmn{kpc})^3$, located immediately downstream of the subcluster, at $t
= 3\ \rmn{Gyr}$ (Table \ref{rms}). The simulations using the novel AMR
criteria are able to get much higher turbulent velocities than run
$A$. These values are in the range predicted theoretically
\citep{ssh06} and found numerically in similar works
\citep{asp03,t05a,t05b}. Interestingly, the new AMR criteria bring
$v_{\rmn{rms}}$ close to the value which is obtained by an equivalent
static grid simulation (cf.~Sect.~\ref{static}). 

A closer analysis of the subcluster tail shows that, at a distance
from the subcluster core $r \approx 5\ r_\rmn{c}$, the density
structure slightly loses contrast. This is an artifact introduced by
the cutoff in the dark matter density and gravitational acceleration
(Sect.~\ref{setup}). We tested that the results are not harmed by this
shortcoming. Moreover, the vorticity pattern is basically not affected
by the cutoff (Figs.~\ref{run-d} and \ref{run-e}) and the same is true
for the new refinement criteria. 

\begin{table}
\caption{Occupation fraction of the AMR levels at $t = 2\ \rmn{Gyr}$,
  for the runs $A$, $D$ and $E$ (the AMR level 0 corresponds to the
  root grid). The fraction is normalised to the whole computational
  domain. For computational efficiency, in all simulations the
  refinement is allowed only in the region of interest around the
  subcluster, with a volume fraction 0.608.} 
\label{occupation}
\centering
\begin{tabular}{c| c c c}

AMR level & run $A$ & run $D$ & run $E$ \\
\hline
1 & 0.150 & 0.156  & 0.422  \\
2 & 0.100 & 0.076  & 0.255  \\
3 & 0.052 & 0.045  & 0.126  \\
4 & 0.027 & 0.031  & 0.074  \\
5 & 0.015 & 0.022  & 0.045  \\
\hline
\end{tabular}
\end{table}

From the point of view of the AMR performance, Table \ref{occupation}
summarises the volume occupation fractions at different levels for
three of the runs under examination, at $t = 2\ \rmn{Gyr}$. As
expected, the simulations $D$ and $E$ have larger occupation fractions
than run $A$. In all cases, the ratio between the occupation fractions
at levels $n$ and $n+1$ lies between 1.4 and 2.0, and only a small
fraction of the computational domain is refined at the higher
levels. Though the overall quality of the AMR operation seems thus
acceptable, a visual inspection of Fig.~\ref{run-d} and \ref{run-e},
right panels, shows that a more efficient use of refinement is made in
run $D$ than in run $E$.  
The latter refines the front bow shock better than the former but,
despite of the implemented refinement cutoff, it triggers some
spurious grids at the sides of the subcluster. This issue is not
severe and is perfectly manageable within the available computational
resources. In a general sense, this is a good example of tuning AMR to
specific problems, finding a difficult equilibrium between a accurate
description of the flow and a convenient use of the tool.

\subsubsection{Comparison with results from the {\sc zeus} solver}
\label{zeus}

It is well known that the simulation of the development of
hydrodynamical instabilities and the onset of turbulence depends
sensitively on the adopted numerical scheme. This issue is extensively
addressed by \citet{ams06}, in a simulation setup which differs from
ours only for the role of gravity, and for the density profile of the
subcluster. A similar, detailed study of our setup is out of the scope
of this paper, but we performed a simulation similar to run $A$ (5 AMR
levels, AMR criterion ``1'') using the {\sc zeus} solver available in {\sc enzo}
(cf.~\citealt{sn92a,sn92b}) instead of PPM. The use of {\sc enzo} with this
solver has been tested by \citet{ams06} in static grid simulations,
whereas here it is applied to an AMR run.  

\begin{figure}
  \resizebox{\hsize}{!}{\includegraphics{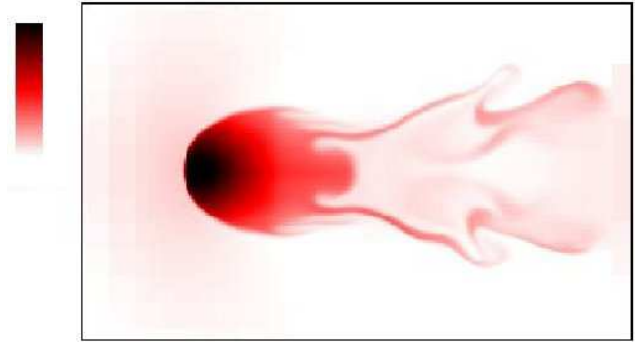}}
  \caption{Density slice of the subcluster at $t = 2\ \rmn{Gyr}$
    (cf.~Fig.~\ref{figure-evolution}, upper right panel), in the
    simulation with the AMR parameters as in run $A$, but using the
    {\sc zeus} solver of {\sc enzo}. The density is linearly colour coded
    following the colour bar on the left.} 
  \label{zeus-dens}
\end{figure}

\begin{figure}
  \resizebox{\hsize}{!}{\includegraphics{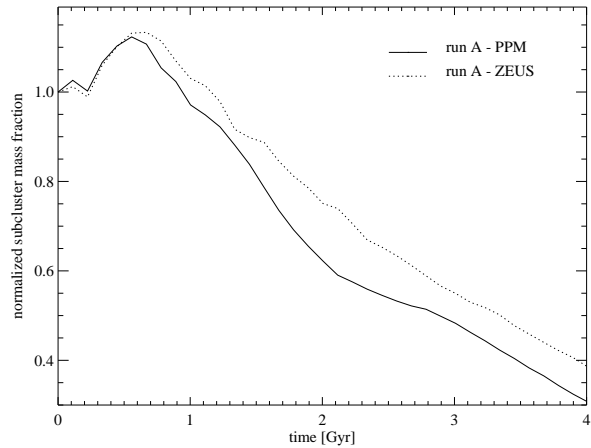}}
  \caption{Temporal evolution of subcluster baryonic mass fraction for
    the runs $A$ (solid line) and the run using the {\sc zeus} solver
    (dotted line). The mass fraction is normalised to its value at $t
    = 0\ \rmn{Gyr}$.} 
  \label{cmf-zeus}
\end{figure}

\begin{figure*}
\centering
  \includegraphics[width=17cm]{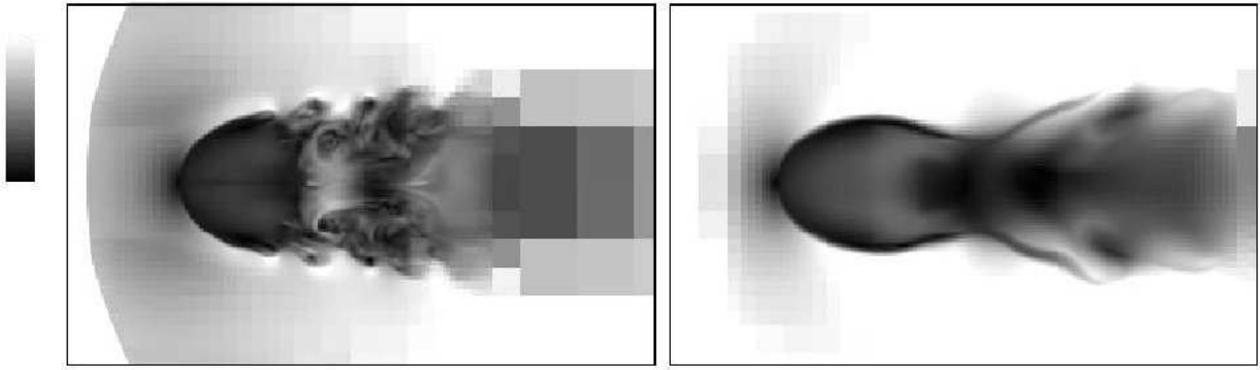}
  \caption{Slices of the subcluster at $t = 2\ \rmn{Gyr}$
    (cf.~Fig.~\ref{figure-evolution}, upper right panel), showing the
    velocity magnitude. Left: run $A$. Right: run using the {\sc zeus}
    solver. The velocity magnitude is linearly coded such that
    that the colour-bar ranges from $v = 0$ (black) to the imposed
    background velocity ($v = 1.6\times 10^3\ \rmn{km\ s^{-1}}$,
    white).} 
  \label{a-vs-zeus}
\end{figure*}

The results of this simulation show clearly that the {\sc zeus} solver is
unable to follow the onset of the KHI and the subsequent development
of turbulent motions in the subcluster wake, where no clear turnover
eddy is visible (Fig.~\ref{zeus-dens}).  
The evolution of the subcluster mass fraction (Fig.~\ref{cmf-zeus}) is
typical of simulations where the turbulent mixing is not efficiently
acting (examples are in some figures of \citealt{ams06}, or in our run
$B$).  Moreover, both the magnitude and the turbulent features of the
flow are not well reproduced in the wake immediately behind the
substructure (compared with run $A$ in Fig.~\ref{a-vs-zeus}).  

The described features do not depend on the chosen artificial
viscosity of the {\sc zeus} solver (set at 2.0 in the performed run;
cf.~\citealt{ams06}). We compared the results with a static run
performed with the {\sc zeus} solver and the same level of resolution of the
AMR run, finding no substantial difference. The relevance of this test
confirms that any conclusion on simulations of turbulent flows must be
critically considered in light of the used numerical scheme, as will
be further discussed in Sect.~\ref{discussion}.

\subsubsection{Comparison with a static grid simulation}
\label{static}

The computational benefit from the use of AMR is especially needed in
simulations where the corresponding static grid is too expensive; this
is often the case in cosmological simulations of structure
formation. On the other hand, most of the presented runs have an
effective resolution of $512^3$, making the comparison with a static
grid simulation still possible. In view of further and computationally
more demanding applications, a comparison of AMR and static grid
simulations is therefore particularly useful. 

\begin{figure*}
\centering
  \includegraphics[width=17cm]{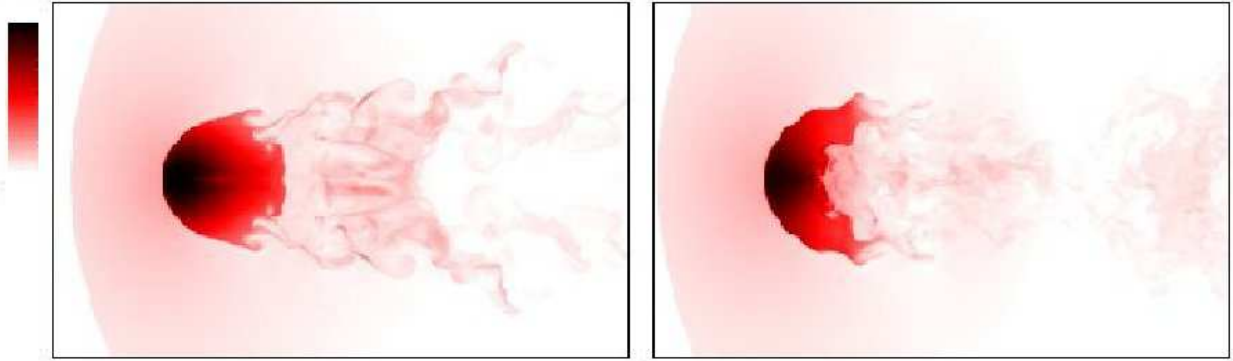}
  \caption{Density slices of the subcluster in the simulation with a
    $512^3$ static grid. Left panel: $t = 2\ \rmn{Gyr}$. Right panel:
    $t = 3\ \rmn{Gyr}$. The density is linearly colour coded
    following the colour bar on the left.} 
  \label{static-dens}
\end{figure*}

A static grid simulation of the subcluster setup has been performed
with a fixed resolution of $512^3$ cells, corresponding to the
effective resolution of the runs $A$, $D$ and $E$. The density slice
at $t = 2\ \rmn{Gyr}$ (Fig.~\ref{static-dens}, left) is similar
to the AMR runs, with only a higher level of symmetric structure in
the wake. The symmetry is not preserved at $t = 3\ \rmn{Gyr}$
(Fig.~\ref{static-dens}, right), when the turbulent tail is very
similar to the runs with new AMR criteria (Figs.~\ref{run-d} and
\ref{run-e}).

\begin{figure}
  \resizebox{\hsize}{!}{\includegraphics{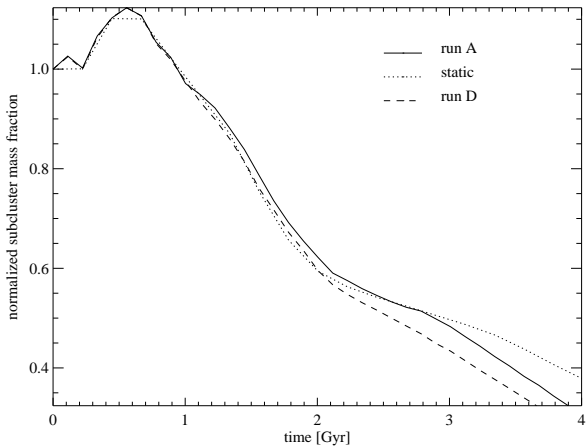}}
  \caption{Temporal evolution of subcluster baryonic mass fraction for
    the runs $A$ (solid line), the run with a $512^3$ static grid
    (dotted line) and the run $D$ as representative of the runs with
    new AMR criteria (dashed line). The mass fraction is normalised to
    its value at $t = 0\ \rmn{Gyr}$.} 
  \label{cmf-full}
\end{figure}

The evolution of the subcluster mass fraction, shown in
Fig.~\ref{cmf-full}, further elucidates the difference between the AMR
and the static grid runs. Until $t = 2\ \rmn{Gyr}$, the static run
shows a slightly larger gas stripping than run $A$, similar to run
$D$, indicating a more effective description of the onset of the KHI
and of the following back-flow. In the late phase and especially after
$t = 3\ \rmn{Gyr}$ this trend reverses, probably because the AMR runs
fail to resolve the finest details of the dispersing structures in
the subcluster tail.  

To sum up, the comparison with the static grid run shows that the new
AMR simulations are able to correctly reproduce the main features of
the subcluster evolution (see also Table \ref{rms}), though the fixed
grid is more effective in the very late phase of the gas
stripping. The benefit of AMR can be fully appreciated also from the
computational point of view: in the case of idealised subcluster
simulations, the use of such a static grid causes an additional
computational cost of a factor from 15 to 40, depending on the AMR
criteria, with increases also for the memory requirements and the
output storage.

\section{Discussion and conclusions}
\label{discussion}

We presented an interesting case of new refinement techniques,
developed for resolving turbulent flows in AMR hydrodynamical
simulations, which may have useful applications in astrophysical
problems. The AMR is not only a tool used for this study, but has 
itself been investigated and tested in its capabilities. New refinement
criteria based on the regional variability of control variables of the
flow were introduced and shown to be superior for simulations
of turbulent media.

As a first application, we presented a set of simulations of an idealised
substructure in a wind in order to study a subcluster
merger in a simplified fashion. In this setup, the novel AMR criteria
are suitable for the study of the turbulent wake, which is better
resolved than using refinement criteria based on local gradients of
selected hydrodynamical variables. The optimal control of the refining
procedure ensures high resolution where desired, as pointed out by the
comparison with an equivalent static grid simulation. 

The overall evolution of the subcluster 
 is similar to the hydrodynamical simulations of \citet{asp03},
 \citet{t05a} and \citet{t05b}, where the development of shear
 instability leads to the formation of a turbulent wake. The typical
 rms velocity in the wake is of the order of $500\ \rmn{km\ s^{-1}}$,
 similar to the above cited simulations and to the theoretical
 predictions \citep{ssh06}. This value is obtained only in runs using
 the new AMR criteria and tends to the result from an equivalent
 static grid simulation, whereas in the reference run $A$ the inferred
 turbulent velocity is smaller by a factor of 2.  

We investigated the proper refining of the turbulent wake with several
numerical experiments. The choice of the AMR criteria has an impact
both on the turbulent velocity and on the morphology of the
subcluster core as a result of the back-flow. We claim that
this issue should be carefully considered in numerical simulations of
subcluster mergers, as for instance in studies of cold fronts. If
the problem is addressed with an inadequate numerical setup
(as in run $A$), the turbulent back-flow will be underestimated
significantly. In extreme cases (run $B$, or Sect.~\ref{zeus}), the
simulation might fail to reproduce the flow
downstream of the subcluster. We further showed that refining
turbulence in the wake does not depend only on AMR but, as expected,
also on the hydrodynamic solver. A detailed code comparison in this
setup, like the one performed for the blob test by \citet{ams06},
would certainly be useful.  

As already stated, we do not intend to discuss the features of the
cold front which forms ahead of the subcluster because our
simulations do not include the magnetic field evolution and/or
heat conduction \citep{afm04,afm05,afm07,xcd07}. The observed cold
front temperature profiles (the best studied example is A3667) call
for the suppression of the thermal conduction at the interface between
the cold subcluster gas and the hot ICM background \citep{mv07}. This
suppression, in turn, would be naturally explained by a magnetic flux
tangential to the cold front surface. \citet{dp07} show, with MHD
simulations using AMR, that the flow past a moving bullet generates
vorticity, even if a magnetic field stabilises its surface against
the KHI. The presence of MHD turbulence and further amplification of
the magnetic field is also suggested. Despite the simplified
approach of both setups, we infer that the generation of a turbulent
wake and a proper modelling of this phase are therefore crucial {\em
  also} for the subcluster MHD simulations, and particular care in
modelling the turbulent flow should be taken also in this class of
simulations (with AMR or not). Only a clear assessment of the
capability of a code in dealing with turbulent flows can allow to
distinguish, for example, between the instability suppression caused
by a magnetic field, and an unreliable modelling of the KHI onset. 

Our simplified approach to the minor merger case shares the 
limitations of previous, similar simulations cited above. However, it
has the advantage of presenting a 
well controlled setup, which would be difficult to study in detail in
the framework of a full cosmological simulation. In view of the
results it is necessary to note that, in the performed simulations,
the merging subcluster was embedded in a background with a very simple
velocity field. In a less idealised flow (cf.~\citealt{nb99,dvb05}), it
is admittedly unlikely that the turbulent tail would extend for
several subcluster radii without being mixed. Nonetheless, it would not
prevent from testing the turbulent character of the flow immediately
downstream of the subcluster \citep{t05b}, both observationally and in
cosmological simulations.  

From an observational point of view, the turbulence level of the
subcluster wake can be an interesting test for supporting the studied
scenario. The rms velocity behind the subcluster can be used as an
observable for turbulence for future X-ray spectrometers. This test
would pertain to the wider, long-standing problem of detecting
turbulence in galaxy clusters \citep[][and reference therein, for an
overview]{sc06}. 

Although significant efforts are undertaken for modelling turbulent
flows in SPH simulations \citep{dvb05,vtc06}, grid-based AMR
simulations are a very powerful approach to simulations in this
field. The AMR criteria used in this work will be further applied to
astrophysical problems, the next step being the simulation of galaxy
cluster formation and evolution (cf.~Paper II). According to the analysis of
\citet{ssh06}, besides the minor merger phase there are several other physical
regimes when turbulence evolves and can play an important role in the
energy budget of a galaxy cluster. The present study corroborates the
importance of the minor mergers in the production of turbulence in
galaxy clusters. Moreover, it provides new tools for the theoretical
study of this problem, which will be continued in Paper II. The final
goal of our investigation is a consistent AMR modelling of turbulent
flows based on a subgrid scale model approach
(cf.~\citealt{snh06,snhr06}). This will be the topic of forthcoming
work.

\section*{acknowledgements}
The numerical simulations were carried out on the SGI Altix 4700 {\it
  HLRB2} of the Leibnitz Computing Centre in Munich (Germany). Thanks go to K.~Dolag for having read the manuscript. The
research of LI and JCN was supported by the Alfried Krupp Prize for
Young University Teachers of the Alfried Krupp von Bohlen und Halbach
Foundation.

\bibliography{cluster-index}
\bibliographystyle{bibtex/mn-web}

\bsp

\label{lastpage}

\end{document}